\documentclass[useAMS,usenatbib]{mn2e}

\usepackage{epsfig}
\usepackage{graphics}
\usepackage{amsmath}
\usepackage[usenames]{color}
\usepackage{amssymb}
\usepackage{amsmath}
\usepackage{times}

\newcommand{\be}{\begin{equation}}
\newcommand{\ee}{\end{equation}}
\newcommand{\bdm}{\begin{displaymath}}
\newcommand{\edm}{\end{displaymath}}
\newcommand{\bea}{\begin{eqnarray}}
\newcommand{\eea}{\end{eqnarray}}

\newcommand{\cf}{\textit{cf.~}}
\newcommand{\ie}{\textit{i.e.,}~}
\newcommand{\eg}{\textit{e.g.,}~}

\newcommand{\Hz}{\,{\rm Hz}\,}

\begin{document}

%%%%%%%%%%%%%%%%%%%%%%%%%%%%%%%%%%%%%%%%%%%%%%%%

\title[QPOs in Bondi-Hoyle accretion]{On the development of QPOs
  in Bondi-Hoyle accretion flows}

\author[O. Donmez,  O. Zanotti and L. Rezzolla]{O. D\"{o}nmez$^{1}$\thanks{E-mail:
odonmez@nigde.edu.tr},  O. Zanotti$^{2}$ and L. Rezzolla$^{2,3}$ \\
$^{1}$Nigde University, Department of Physics, Nigde Turkey, 51200 \\
$^{2}$Max-Planck-Institut f{\"u}r Gravitationsphysik, Albert Einstein Institut, Golm, Germany \\
$^{3}$Department of Physics, Louisiana State University, Baton Rouge, LA USA}

\date{}

\pagerange{\pageref{firstpage}--\pageref{lastpage}} \pubyear{2010}

\maketitle

\label{firstpage}

\begin{abstract}
The numerical investigation of Bondi-Hoyle accretion onto a moving
black hole has a long history, both in Newtonian and in
general-relativistic physics. By performing new two-dimensional and
general-relativistic simulations onto a rotating black hole, we point
out a novel feature, namely, that quasi-periodic oscillations (QPOs)
are naturally produced in the shock cone that develops in the
downstream part of the flow. Because the shock cone in the downstream
part of the flow acts as a cavity trapping pressure perturbations,
modes with frequencies in the integer ratios $2:1$ and $3:1$ are
easily produced. The frequencies of these modes depend on the
black-hole spin and on the properties of the flow, and scale linearly
with the inverse of the black-hole mass. Our results may be relevant
for explaining the detection of QPOs in Sagittarius A$^\ast$, once
such detection is confirmed by further observations. Finally, we
report on the development of the flip-flop instability, which can
affect the shock cone under suitable conditions; such an instability
has been discussed before in Newtonian simulations but was never found
in a relativistic regime.
\end{abstract}


%###############################################################################
%###############################################################################

\section{INTRODUCTION}
\label{Introduction}

Non-spherical accretion flows are astrophysically relevant in all those
situations where strong winds with a small amount of angular momentum
are able to transport considerable amounts of matter towards an
accreting compact object, \eg a black hole or a neutron star.
Examples include the case of massive X-ray binaries in which a compact
object accretes from the wind of an early-type star, or the case of
young stellar systems orbiting in the gravitational potential of their
birth cluster and accreting from the dense molecular interstellar
medium. One of the most celebrated and best studied types of
non-spherical accretion flows is the
Bondi-Hoyle~\citep{Hoyle1939,Bondi1944} which, we recall, develops
when a black hole moves relative to a uniform gas cloud. The
Bondi-Hoyle flow has been the subject of several numerical
investigations, starting from~\citet{Matsuda1987} and followed
by~\citet{Fryxell1988, Sawada1989, Benensohn1997}. As a summary of the
bulk of work done so far on this topic, Table 1
of~\citet{Foglizzo2005} reports an overview of published numerical
simulations of Bondi-Hoyle-Lyttleton accretion over the last $30$
years and lists more than $40$ works. While providing detailed
descriptions of the morphology of the gas capture in the supersonic
regime, several of the above mentioned investigations were aimed at
verifying the occurrence of the so called flip-flop instability. This
consists of an instability to tangential velocities of the shock cone
that forms in the downstream region of the flow and it manifests in
the oscillation of the shock cone from one side to the other of the
accretor.

In the relativistic regime, the first two-dimensional simulations of
Bondi-Hoyle accretion were performed by~\citet{Petrich89} and
subsequently by~\citet{Font98a,Font1998g,Font1999b}. These
investigations considered flows both in axisymmetry and in the
equatorial plane; interestingly none of these works showed the
occurrence of the flip-flop instability, so that no evidence existed,
prior to this work, about the development of the instability also in a
curved spacetime. More recently, a fully general relativistic
investigation of the Bondi-Hoyle accretion has been considered by
\citet{Farris2010} in the context of the merger of supermassive
black-hole binaries; in this case, however, no discussion of the
flip-flop instability was presented.

However, the occurrence of the instability is not the only relevant
physical process that may manifest in Bondi-Hoyle accretion flows. An
additional one, that does not seem to have been considered so far, is
related to the possibility that the shock cone traps oscillation
modes, thus producing Quasi Periodic Oscillations (QPOs). The
intuition is indeed simple yet rather suggestive. The typical shocked
cone of the Bondi-Hoyle accretion is likely to provide a natural
cavity for the development and confinement of oscillation modes of
sonic nature. If the accretor is a compact object, Newtonian
hydrodynamics is a good approximation only at large radial distances,
while it can lead to misleading conclusions when studying the flow
evolution close to the event horizon. In this paper we perform two
dimensional general relativistic hydrodynamics simulations by focusing
on the dynamical behavior of the Bondi-Hoyle shock cone under a wide
range of parameters. The QPOs that we have found have typical
frequencies in the range from $10^{-5}$\Hz to $10^{-3}$\Hz for a black
hole with $M=10^6M_{\odot}$ (and in the range from $1$\Hz to $10^2$\Hz
for a black hole with $M=10\,M_{\odot}$). As a result, they can become
relevant for the interpretation of QPOs observed both in the galactic
center and in high-mass X-ray binaries. Our analysis has also shown
that the flip-flop instability does occur even in the relativistic
framework, and we have investigated how such instability can interplay
with the QPO phenomenon, by suppressing or exciting specific modes of
oscillations.

The plan of the paper is the following: in
Section~\ref{Mathemtical_formulation} we provide an overview of the
hydrodynamical equations, of the numerical methods adopted and of the
physical set up of the problem. In Sec.~\ref{Results} we describe our
results, while Sec.~\ref{Astrophysical_applications} contains a
discussion of the potential applicability to two specific
astrophysical cases. Finally, Sec.~\ref{Discussion_and_conclusions}
is devoted to the conclusions of our work. We use a geometrized
system of units with $G=c=1$.

%###############################################################################
%###############################################################################
%###############################################################################
%###############################################################################

\section{Mathematical formulation}
\label{Mathemtical_formulation}

\subsection{General relativistic hydrodynamics equations}

We study Bondi-Hoyle accretion of a perfect fluid in the curved
background spacetime of a rotating black hole. The energy momentum
tensor of the fluid has the usual form
\be
T^{\mu\nu}= h\rho u^\mu u^\nu + p g^{\mu\nu} \,,
\ee
where $u^\mu$ is the four velocity of the fluid, $h$, $\rho$ and $p$
are the specific enthalpy, the rest-mass density and the pressure,
respectively. All of these quantities are measured by an observer
comoving with $u^\mu$, while $g^{\mu\nu}$ is the metric of the
spacetime. By adopting the $3+1$ formalism of general relativity and
after choosing Boyer-Lindquist coordinates $(t,r,\theta,\phi)$, the
line element of the metric is written as
\begin{eqnarray}
ds^2 &=&-\left(1-\frac{2Mr}{\Sigma^2}\right)dt^2 - \frac{4Mar}{\Sigma^2}\sin^2\theta
dtd\phi \nonumber \\ 
&&+\frac{\Sigma^2}{\Delta}dr^2 + \Sigma^2 d\theta^2 \nonumber
+\frac{A}{\Sigma^2}\sin^2\theta d\phi^2 \,,
\label{line element1}
\end{eqnarray}
where 
\bea
\Sigma^2 &\equiv& r^2 + a^2 \cos^2(\theta)\,, \\
\Delta &\equiv& r^2 - 2Mr + a^2 \,,\\
A&\equiv&(r^2+a^2)^2-a^2\Delta \sin^2\theta\,,
\eea
with $a$ and $M$ being the spin and the mass of the black hole. The
{\em lapse} function and the {\em shift} vector of the metric are
given, respectively, by $\alpha=(\Sigma^2\Delta/A)^{1/2}$ and
$\beta^i=(0,0,-2Mar/A)$. When solving numerically the
general-relativistic hydrodynamic equations it is important to write
them in a conservation form~\citep{Banyuls97}
\begin{equation}
\frac{\partial {\bf{U}}}{ \partial t}+ \frac{\partial
  {\bf{F}^i}}{\partial q^i} = {\bf{S}}\,.
\label{hydro1}
\end{equation}
where ${\bf{U}}$, ${\bf{F}}^i$ and ${\bf{S}}$ are the vectors of the
conserved variables, of the fluxes and of the sources, respectively,
while $q^i$ is the generalized coordinate in the $i-$th
direction. When performing two-dimensional numerical simulations in
the equatorial plane, Eq.\eqref{hydro1} reduces to
\begin{equation}
\frac{\partial {\bf{U}}}{\partial t} + \frac{\partial {\bf{F}}^r}{\partial r} + 
\frac{\partial {\bf{F}}^{\phi}}{\partial \phi}= {\bf{S}}\,.
\label{hydro2}
\end{equation}
\noindent
The conservative variables, written in terms of the primitive
variables $(\rho,v^i,p)$, are
\begin{eqnarray}
{\mathbf U} & =& \left[\begin{array}{c}
D \\
S_r \\
S_{\phi}\\
\tau
\end{array}\right]=
\left[\begin{array}{c}
\sqrt{\gamma}W\rho   \\
\sqrt{\gamma}\rho h W^2 V_r   \\
\sqrt{\gamma}\rho h W^2 V_{\phi}  \\ 
\sqrt{\gamma}(\rho h W^2 -p -W\rho)
\end{array}\right]\,,
\label{hydro3}
\end{eqnarray}
where $V^i=u^i/W + \beta^i/\alpha$ is a spatial vector whose indices
are raised and lowered through the spatial metric $\gamma_{ij}$ and it
represents the three-velocity of the fluid with respect to the
Eulerian observer associated to the $3+1$ splitting of the metric. $W
= (1-\gamma_{ij}V^iV^j)^{-1/2}$ is the Lorentz factor of the fluid and
$\gamma={\rm det} (\gamma_{ij})=\Sigma^4\sin^2\theta /\alpha^2$ is the
determinant of three metric. Although~\citet{Font1999b} already
reported fluxes and sources for the hydrodynamical equations in the
Kerr metric, we explicitly list them here for convenience as
\begin{eqnarray}
{\mathbf F}^r & =& \left[\begin{array}{c}
\alpha (V^r - \beta^r/\alpha) D \\
\alpha \lbrace (V^r - \beta^r/\alpha) S_r +
\sqrt{\gamma} p \rbrace  \\ 
\alpha (V^r - \beta^r/\alpha) S_{\phi}  \\
\alpha \lbrace (V^r - \beta^r/\alpha) \tau +
\sqrt{\gamma}V^r p \rbrace  
\end{array}\right]\,,
\label{hydro4a}
\end{eqnarray}

\begin{table}
  \caption{Initial models adopted in numerical simulation. From left
    to right the columns report: the name of the model, 
    the black hole spin $a$, the
    asymptotic flow velocity $V_{\infty}$, the asymptotic Mach number
    ${\cal M}_{\infty}$ and the inner boundary $r_{{\rm in}}$ of the
    grid. The final simulation time is set equal to $10000M$ in all of
    the models. The reference adiabatic index adopted is $\Gamma=4/3$,
    though we have also considered different values to explore the
    effect on the results.
 \label{table:Initial Models}}
\begin{center}
  \begin{tabular}{ccccc}
    \hline \hline
   Model & $a/M$ & $V_{\infty}$ &
   ${\cal M}_{\infty}$& $r_{{\rm in}}(M)$ \\
    \hline
%$A$   &  $0.99$& $0.5$   & $5$    & $1.78$  \\
%$B_1$ &  $0.9$ & $0.0001$& $0.001$& $1.78$  \\
$\mathtt{A1}$ &  $0.9$ & $0.001$ & $0.01 $& $1.78$  \\
$\mathtt{A2}$ &  $0.9$ & $0.1$   & $1$    & $1.78$  \\
$\mathtt{A3}$ &  $0.9$ & $0.2$   & $2$    & $1.78$  \\
$\mathtt{A4}$ &  $0.9$ & $0.3$   & $3$    & $1.78$   \\
%$B_6$ &  $0.9$ & $0.35$  & $3.5$  & $1.78$  \\
$\mathtt{A5}$ &  $0.9$ & $0.4$   & $4$    & $1.78$ \\
%$B_8$ &  $0.9$ & $0.45$  & $4.5$  & $1.78$ \\
$\mathtt{A6}$ &  $0.9$ & $0.5$   & $5$    & $1.78$   \\
$\mathtt{A7}$&   $0.9$ & $0.6$   & $6$    & $1.78$   \\
    \hline
%$C$   & $0.8$ & $0.1$   & $1$    & $1.78$   \\
%$D_1$ & $0.5$ & $0.1$   & $1$    & $2.1$  \\
$\mathtt{B}$ &    $0.5$ & $0.4$   & $4$    & $2.1$  \\
%$D_3$ & $0.5$ & $0.5$   & $5$    & $2.1$  \\
%$E_1$ & $0.3$ & $0.5$   & $5$    & $2.1$  \\
%$E_2$ &  $0.3$ & $0.4$   & $4$    & $2.1$  \\
    \hline
$\mathtt{C1}$ &  $0.0$ & $0.001$ & $0.01$ & $2.1$  \\
$\mathtt{C2}$ &  $0.0$ & $0.1$   & $1$    & $2.1$  \\
$\mathtt{C3}$ &  $0.0$ & $0.2$   & $2$    & $2.1$ \\
$\mathtt{C4}$ &  $0.0$ & $0.3$   & $3$    & $2.1$ \\
$\mathtt{C5}$ &  $0.0$ & $0.4$   & $4$    & $2.1$ \\
$\mathtt{C6}$ &  $0.0$ & $0.5$   & $5$    & $2.1$  \\
$\mathtt{C7}$ &  $0.0$ & $0.6$   & $6$    & $2.1$ \\
    \hline
    \hline
  \end{tabular}
\end{center}
%  \tablenotetext{}{}
%\vskip -0.8truecm
\end{table}

\begin{eqnarray}
{\mathbf F}^\phi & =& \left[\begin{array}{c}
\alpha (V^\phi - \beta^\phi/\alpha) D \\
\alpha (V^{\phi} - \beta^{\phi}/\alpha) S_r  \\
\alpha \lbrace (V^{\phi} - \beta^{\phi}/\alpha) S_{\phi}
+ \sqrt{\gamma} p \rbrace  \\ 
\alpha \lbrace (V^\phi - \beta^\phi/\alpha) \tau +
\sqrt{\gamma}V^\phi p \rbrace  
\end{array}\right]
\label{hydro4b}
\end{eqnarray}
and 
\begin{eqnarray}
{\mathbf S} & =& \left[\begin{array}{c}
0 \\
\alpha\sqrt{\gamma}T^{\mu\nu}g_{\nu\sigma}\Gamma^{\sigma}_{\mu
  r} \\
\alpha\sqrt{\gamma}T^{\mu\nu}g_{\nu\sigma}\Gamma^{\sigma}_{\mu
  \phi}\\
\alpha\sqrt{\gamma}(T^{r t}\partial_{r}\alpha - 
\alpha (T^{rt}\Gamma^{t}_{rt} + T^{rr}\Gamma^{t}_{rr} +T^{r\phi}\Gamma^{t}_{r\phi} )) 
\end{array}\right] \nonumber \\
&&
\label{hydro6}
\end{eqnarray}
\begin{figure*}
\psfig{file=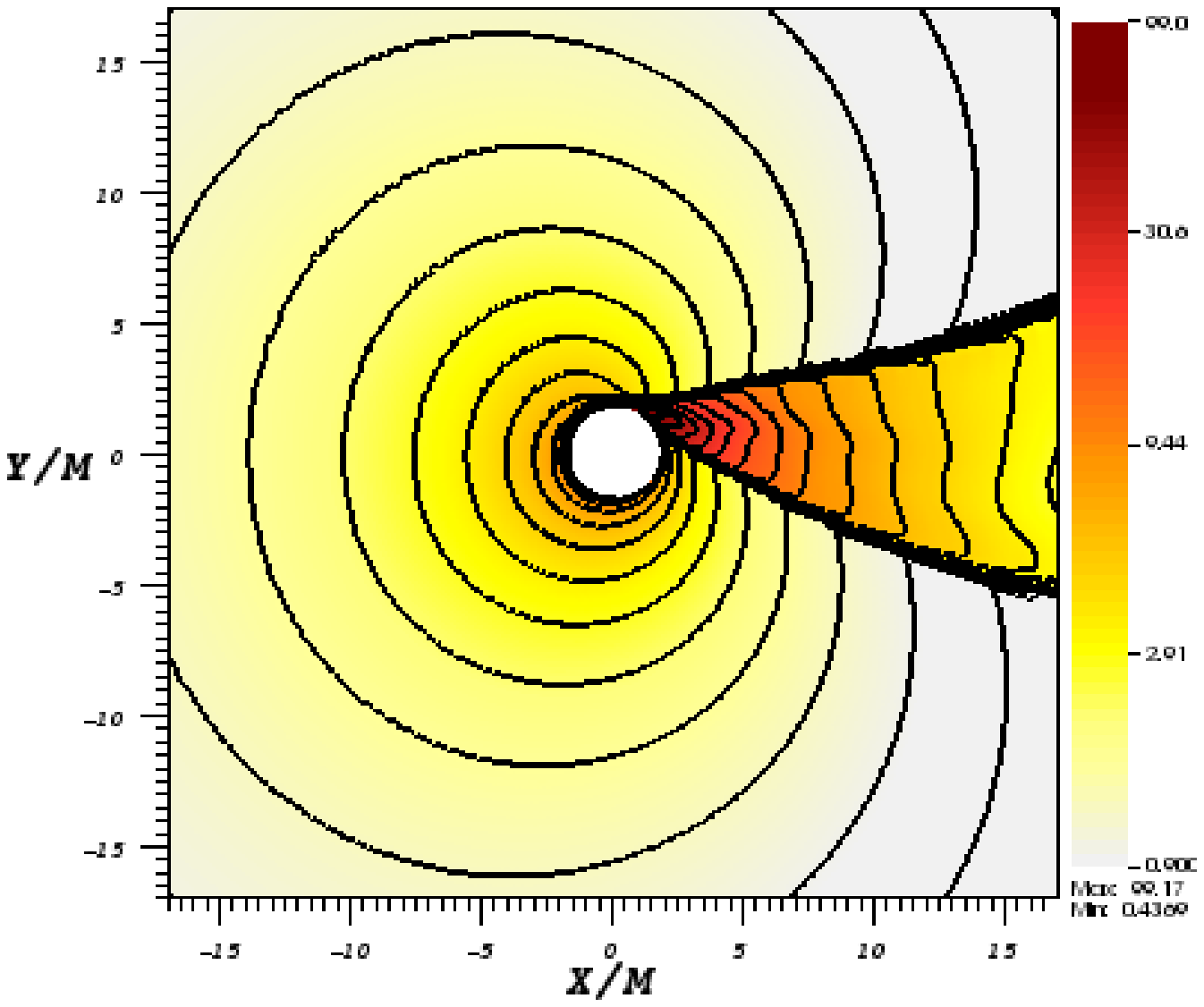,width=8.2cm}
\hskip 1.0cm
\psfig{file=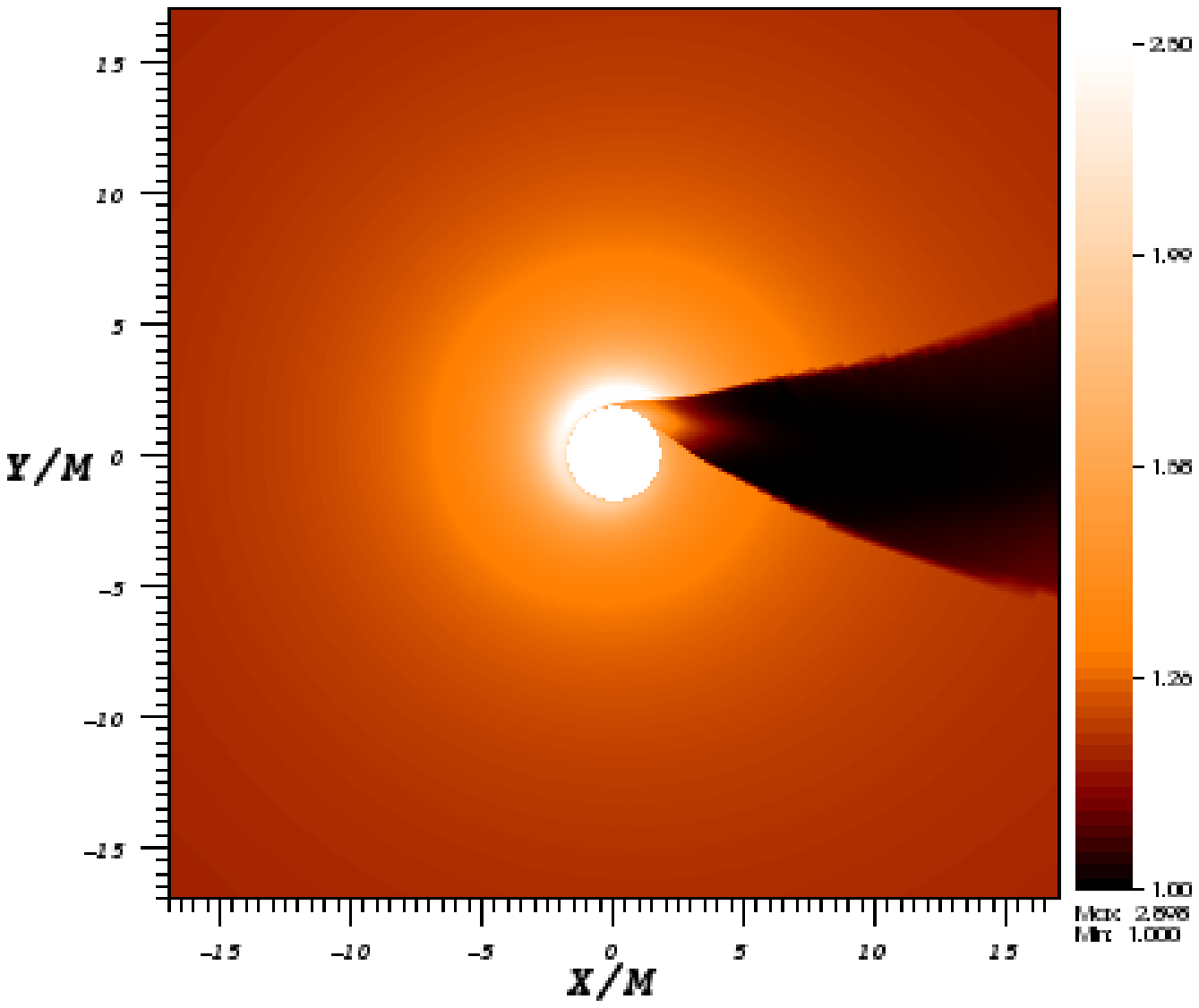,width=8.2cm}
\caption{Rest-mass density distribution (left panel) and Lorentz
  factor distribution (right panel) for a black hole with spin $a=0.9$
  and a supersonic flow with asymptotic velocity $V_{\infty}=0.5$, and
  at time $t=10000\,M$. In the downstream region a clear shock cone
  forms which is distorted by the non vanishing spin of the black
  hole.
\label{fig_2D_stationary}}
\end{figure*}
The equation of state that we adopt is that of an ideal gas, namely
\begin{equation}
 p = (\Gamma -1)\rho \epsilon\,,
\label{EoS}
\end{equation}
where $\Gamma$ and $\epsilon$ are the adiabatic index and the specific
internal energy, respectively. The reference adiabatic index used
  in the simulations is $\Gamma=4/3$, though we have also considered
  different values to explore the effect on the results.

We have solved the system of equations \eqref{hydro1} through
high-resolution shock-capturing schemes (HRSC) based on approximate
Riemann solvers. A {\em minmod} linear algorithm for the
reconstruction of the left and right states at each interface between
adjacent numerical cells is adopted, while the numerical fluxes are
computed with the Marquina flux formula. More specific details about
the numerical scheme can be found in~\citet{Donmez2004}.

As a final remark we note that because we are not here considering any
contribution coming from the cooling of the shocked gas, the
oscillations we will discuss are not related to those reported
by~\citet{Molteni1996, Chakrabarti2004, Okuda2007}. In those studies,
in fact, the presence of cooling processes is a necessary ingredient
for the appearance of oscillations along the standing shocks.

%########################################################################

\subsection{Initial Conditions, Boundary Conditions and Approximations}
\label{Boundary Condition and Numerical Isssues}

\begin{figure*}
\psfig{file=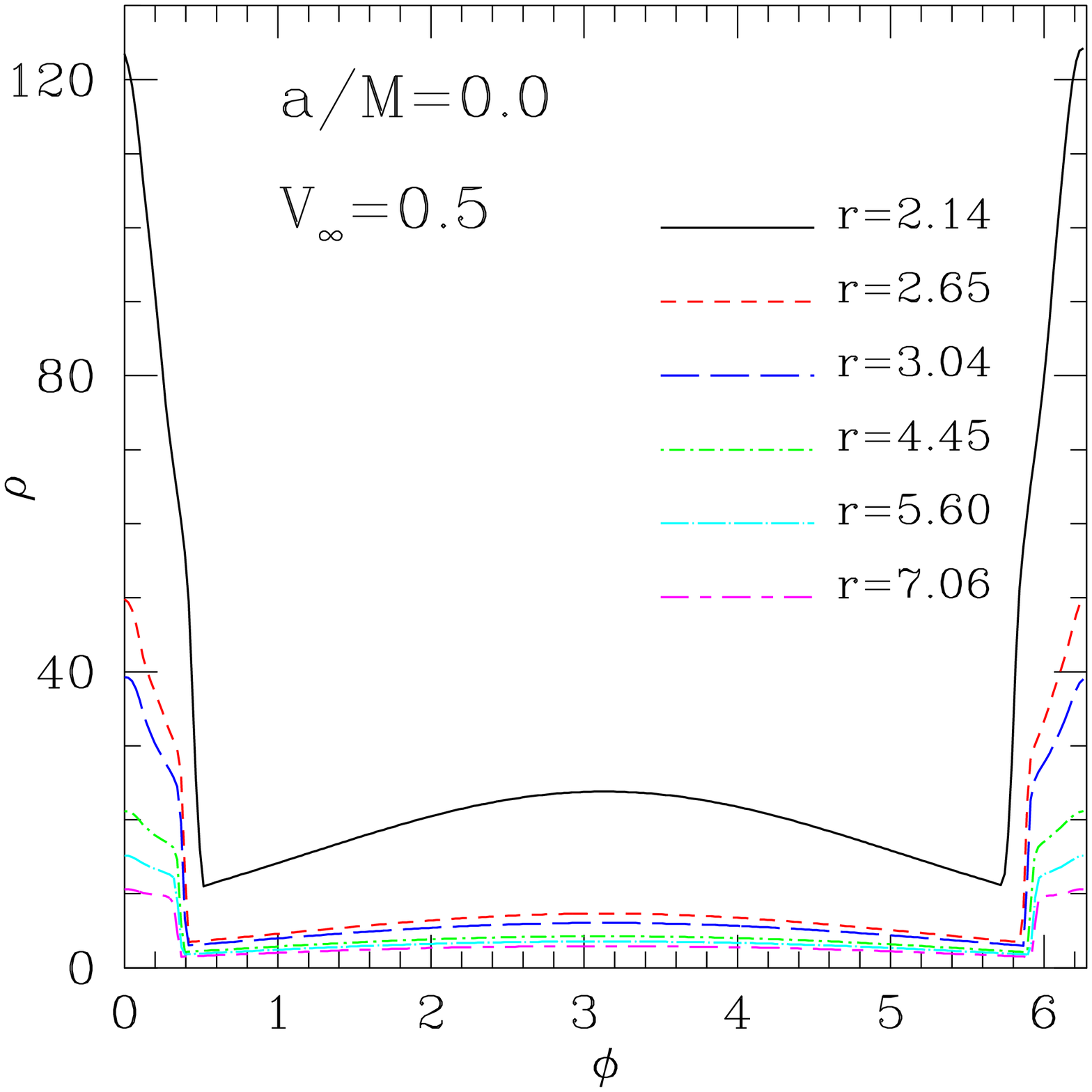,width=8.0cm}
\hskip 1.0cm
\psfig{file=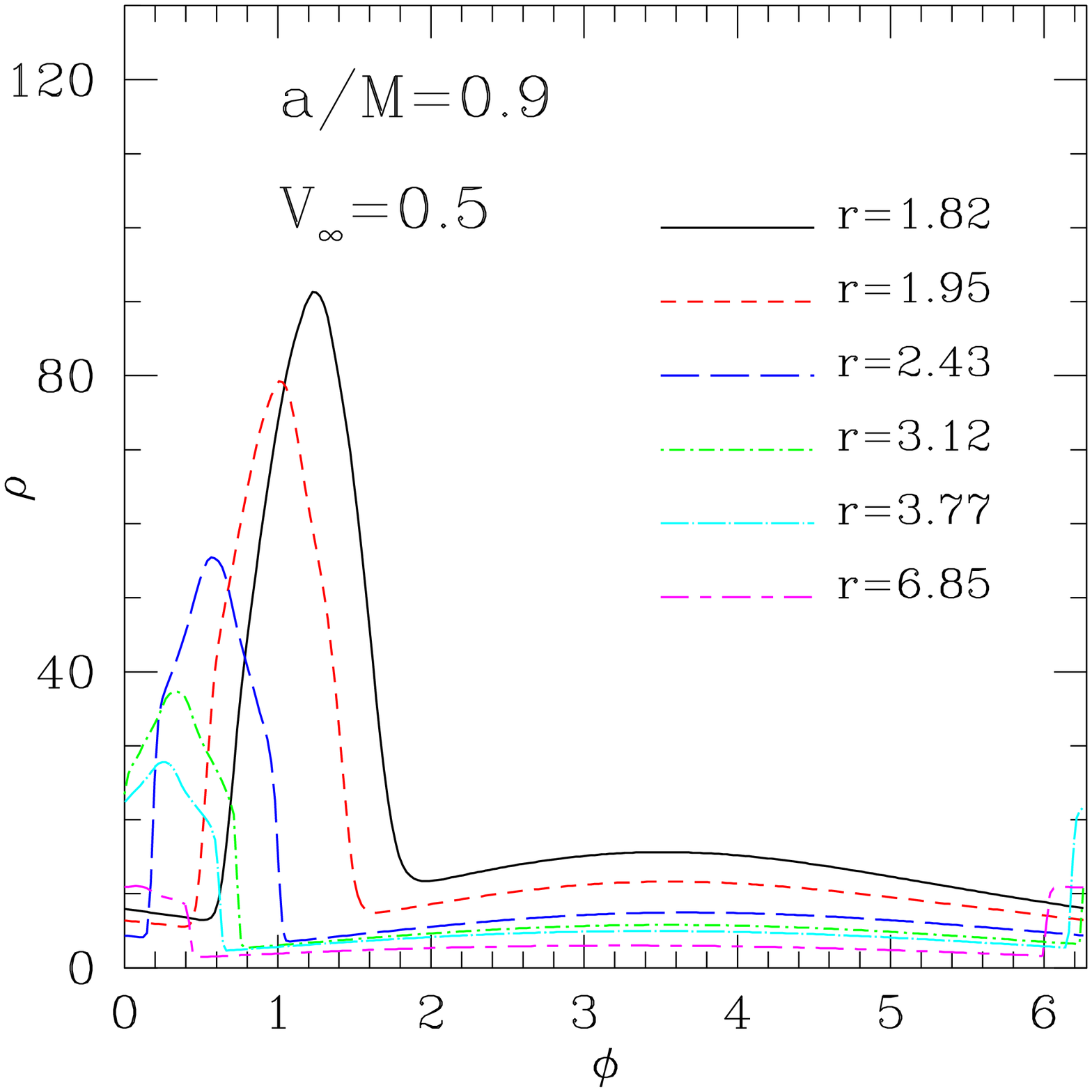,width=8.0cm}
\caption{One dimensional profiles of the rest-mass density at time
  $t=10000\,M$, showing the location of the shock at different radial
  shells for an asymptotic flow velocity $V_{\infty}=0.5$. The left
  panel refers to Schwarzschild black hole, while the right one to a
  rotating black hole with spin $a/M=0.9$. Note that in this latter
  case the shock is distorted and no longer symmetric.}
\label{fig_shock_location}
\end{figure*}

We perform numerical simulations on the equatorial plane, \ie $\theta
= \pi/2$.
The initial velocity field is given in terms of an asymptotic velocity
$V_{\infty}$ as in~\citet{Font1998g}, \ie
\begin{eqnarray}
\label{asymtotic_vel_a}
V^r &=& \sqrt{\gamma^{rr}} V_{\infty} \cos(\phi) \,,\\
V^{\phi} &=&  -\sqrt{\gamma^{\phi \phi}}  V_{\infty} \sin(\phi) \,.
\label{asymtotic_vel_b}
\end{eqnarray}
These relations guarantee that the velocity of the injected gas at
infinity is parallel to the $x-$direction, while $V^2 \equiv V_iV^i=
V_{\infty}^2$ everywhere in the flow. The value of $V_{\infty}$ can
then be chosen to investigate the different regimes of the flow and
to consider therefore subsonic or supersonic accretion. 

During the evolution additional matter is injected supersonically from
the outer boundary in the upstream region with the same analytic
prescription of~\eqref{asymtotic_vel_a} and~\eqref{asymtotic_vel_b},
thus reproducing a continuous wind at large distances. The initial
density and pressure profiles are adjusted to make the sound speed
equal to a required value, which we choose to be
$c_{s,\infty}=0.1$. In practice, once $c_{s,\infty}$ is chosen, we set
the density to be a constant (\ie $\rho=1$) and then the pressure is
derived from the relativistic definition of the sound speed, \ie $p =
c^2_{s,\infty}\rho(\Gamma - 1)/(\Gamma (\Gamma -
1)-c^2_{s,\infty}\Gamma)$. A brief description of the initial models
is reported in Table~\ref{table:Initial Models}, where models
$\mathtt{C*}$ refer to Schwarzschild black holes and different fluid
injection velocities, while models $\mathtt{A*}$ and $\mathrm{B}$
refer to Kerr black holes respectively with spin $a/M=0.9$ and
$a/M=0.5$, again spanning different fluid injection velocities.

The computational grid consists of $N_r \times N_{\phi}$
uniformly spaced zones in the
radial and angular directions, respectively, covering a computational
domain extending from $r_{\rm {min}}$ (reported in
Table~\ref{table:Initial Models}) to $r_{\rm {max}}=43\,M$ and from $0$
to $2\pi$. For our fiducial simulation we have chosen $N_r=512$ and
$N_\phi=256$, but have also verified that the qualitative results (\ie
the appearance of the QPOs or of the instability) are not
sensitive to the resolution used or to the location of the outer
boundary, which has been moved to $r_{\rm max} \simeq 80\,M$ in some tests. 

The boundary conditions in the radial direction are such that at the
inner radial grid point we implement outflow boundary conditions by a
simple zeroth-order extrapolation (\ie a direct copy) of all
variables. At the outer radial boundary, on the other hand, we must
distinguish between the upstream region, with $\pi/2<\phi<3/2\pi$, and
the downstream region, with $-\pi/2<\phi<\pi/2$. In the upstream
region we continuously inject matter with the initial velocity field,
while in the downstream region we use outflow boundary conditions.
Finally, periodic boundary conditions are adopted along the $\phi$
direction.

Finally, we note that in order to validate the code, we have
  carried out a number of comparisons with the \texttt{ECHO} code
  presented in~\citet{DelZanna2007}, which, among other differences,
  uses a non-uniform grid structure. The results obtained from the two
  codes have not shown significant differences.

%###############################################################################
%###############################################################################
%###############################################################################
%###############################################################################

\section{NUMERICAL RESULTS}
\label{Results}
%
%------------------------------------------------------
\subsection{Properties of the stationary pattern}
\label{Properties of the stationary pattern}
%------------------------------------------------------

%
\begin{figure*}
\psfig{file=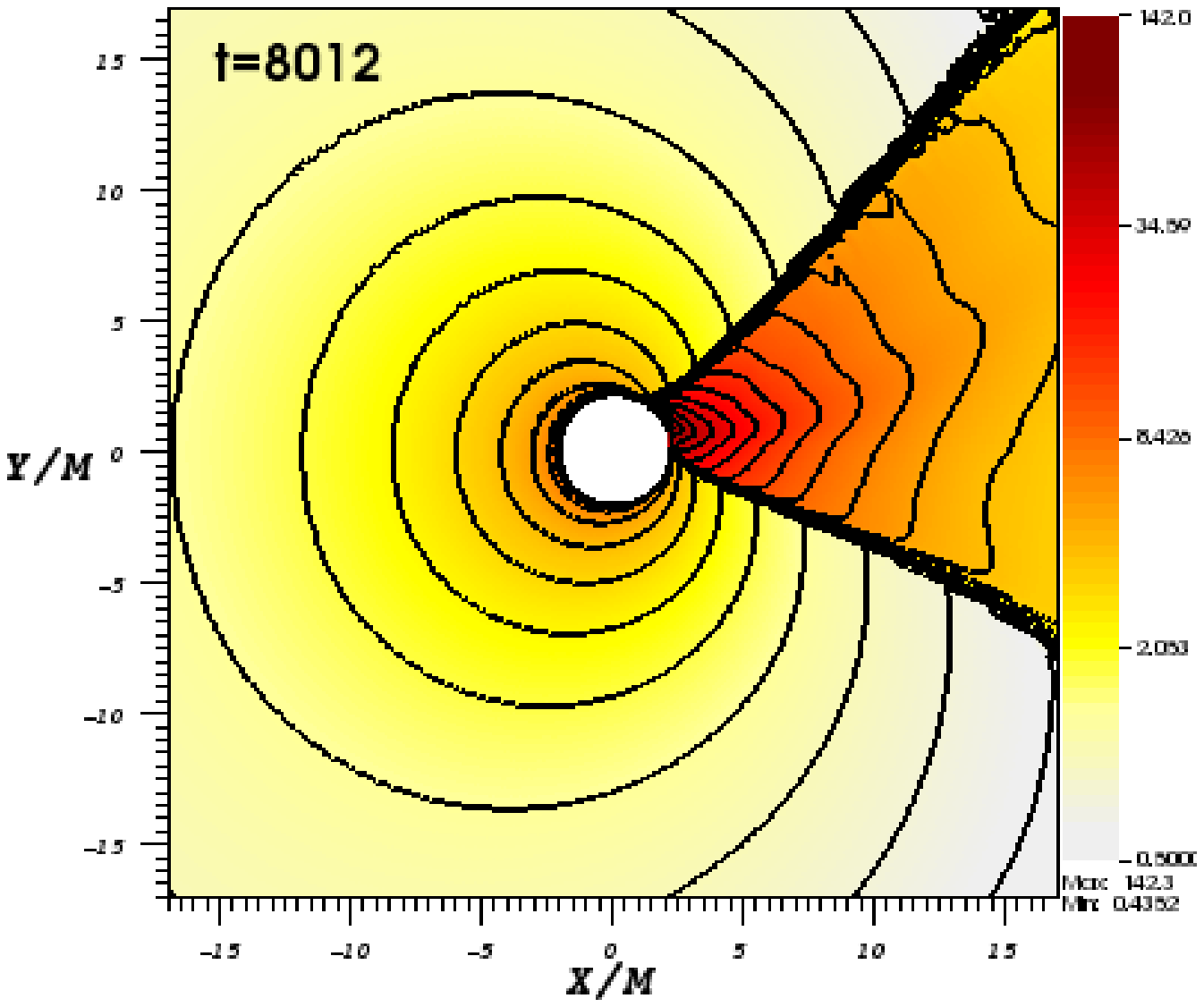,width=8.2cm}
\hskip 1.0cm
\psfig{file=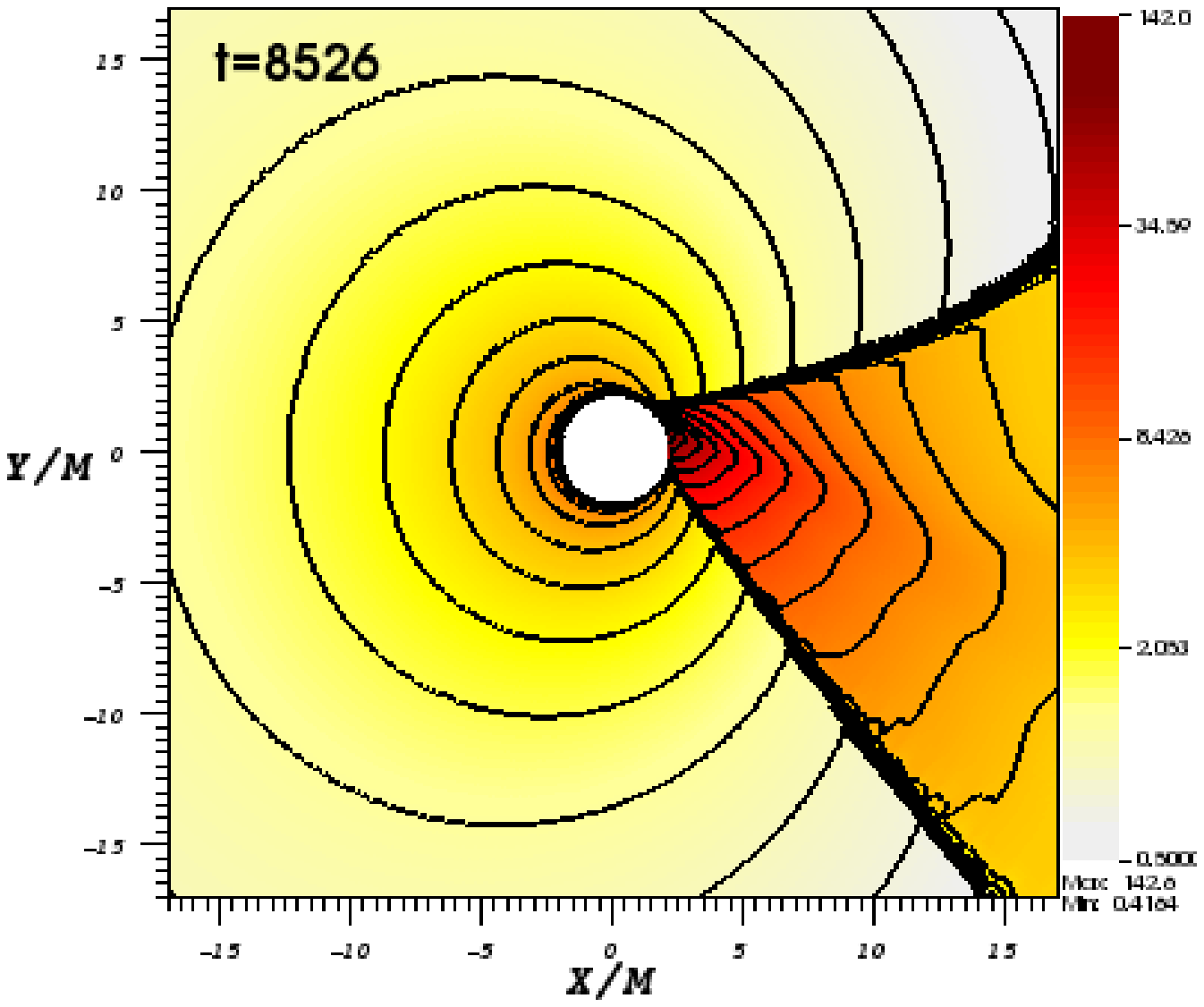,width=8.2cm}
\psfig{file=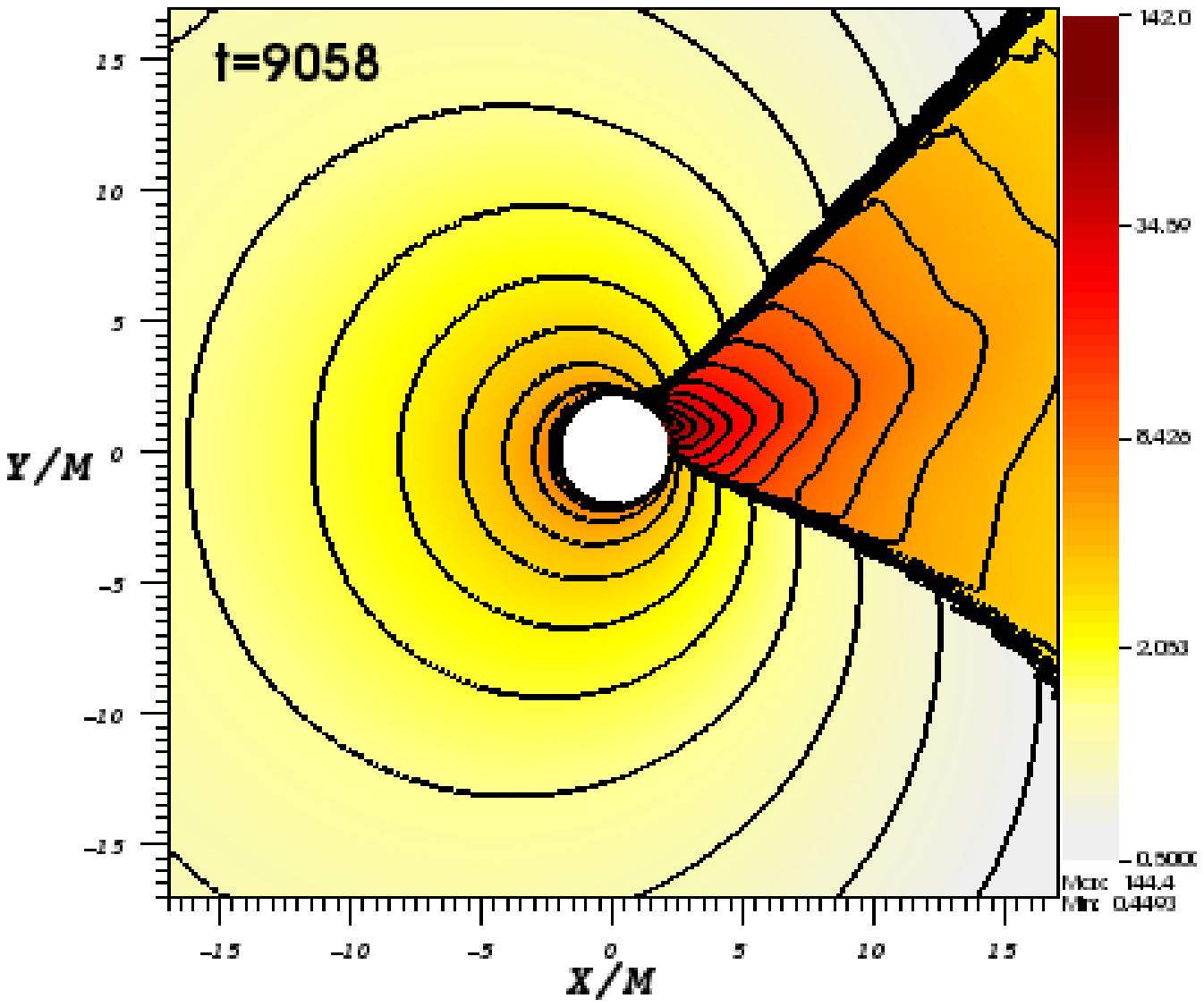,width=8.2cm}
\hskip 1.0cm
\psfig{file=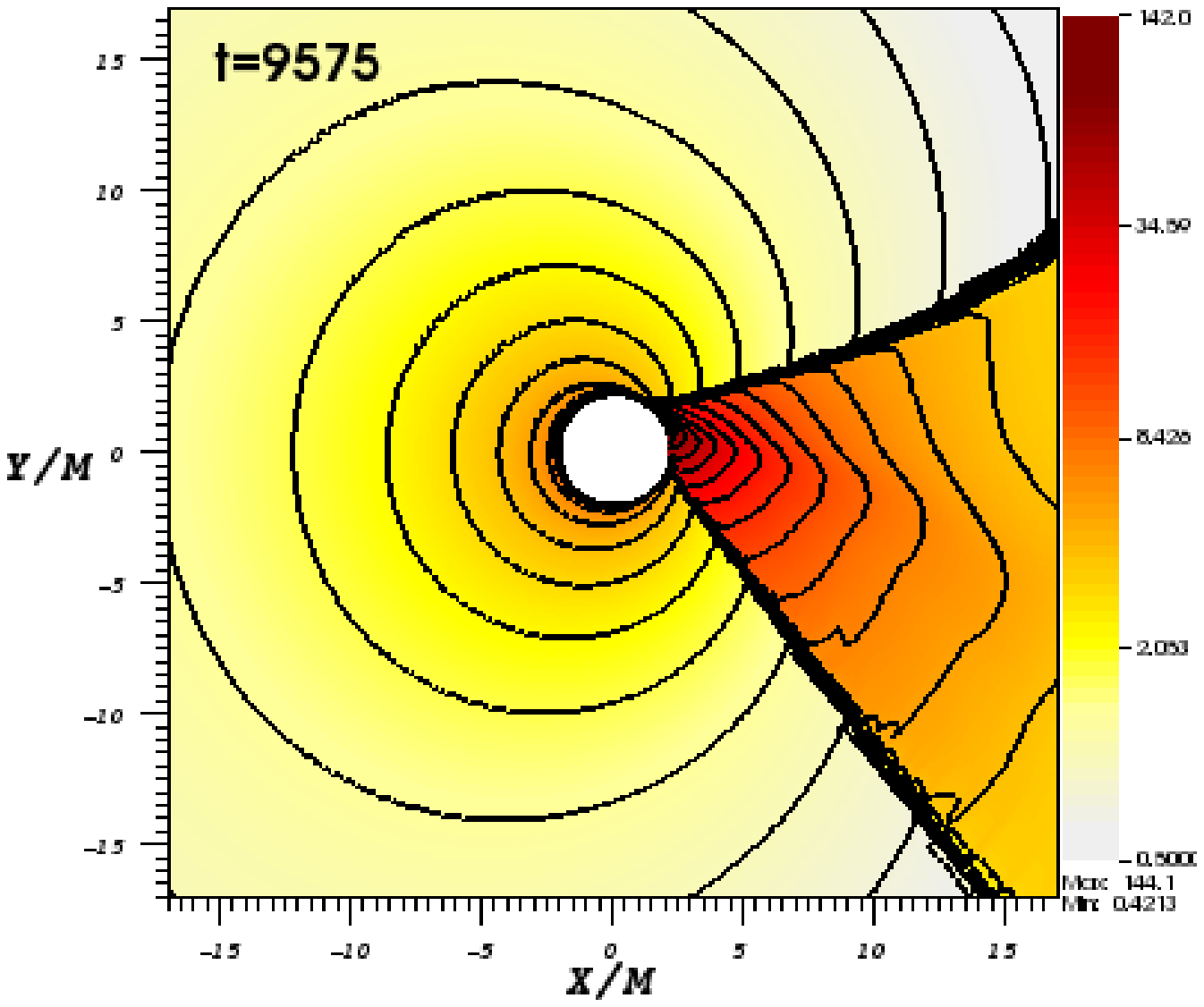,width=8.2cm}
\caption{Logarithm of the density at different times for model
  \texttt{B} with $a/M=0.5$ and $V_{\infty}=0.4$. The flip-flop
  instability manifests after the shock cone has reached a stationary
  state and it causes the shock cone to oscillate back and forth with
  an oscillation period around $1046M$. }
\label{flip-flop}
\end{figure*}

The main features of Bondi-Hoyle accretion were first investigated in
the relativistic framework through a number of numerical simulations
by~\citet{Petrich89,Font98a,Font1998g,Font1999b}. The overall results
obtained can be summarized as follows.

As already shown by several authors, when a homogeneous flow of matter
moves non radially towards a compact object, a shock wave will form in
the neighborhood of the accretor. Depending then on the properties of
the flow, namely on the adiabatic index and on the asymptotic Mach
number ${\cal M}_{\infty}$, the shock can either reach very close to
the accretor or be at a certain distance from it [see,
  \eg~\citet{Foglizzo2005}]. More specifically, for any given value of
${\cal M}_{\infty}$, there is a critical adiabatic index $\Gamma
\simeq 2$, below which a shock wave of conic shape, \ie a ``shock
cone'', forms in the downstream part of the flow, \ie downstream of
the accretor. Matter from the upstream region crosses continuously the
shock front, it undergoes a strong deceleration and, if it is inside
the accretion radius, is ultimately accreted. As a result, the maximum
rest-mass density in the downstream region is always larger than the
corresponding one in the upstream region and, consequently, the mass
accretion rate is significantly non-spherical and larger in the
downstream part of the flow. Interestingly, when suitable coordinates
are used (\eg Kerr-Schild coordinates), it has been shown
by~\citet{Font1999b} that the shock even penetrates the event horizon.
On the other hand, for increasingly stiff fluids and thus for values
of the adiabatic index larger than the critical one, a bow shock wave
forms in the upstream part of the flow, \ie upstream of the
accretor. In this case the accretion is almost spherical and because
the bow shock does not reach into the black-hole's horizon, it is
often referred to as a ``detached'' shock.

The shape of the shock cone depends sensitively on the spin of the
black hole. If the black hole is not spinning, then the shock cone
is perfectly symmetric about the $\phi=0$ direction. On the other
hand, if the black hole is spinning, the induced frame-dragging effect
produces a ``wrapping'' of the shock cone, as already shown and
discussed in great detail by~\citet{Font1999b}. 

Our simulations essentially confirm this picture, and the main
features of the stationary pattern are summarized in
Figure~\ref{fig_2D_stationary}, which shows the two dimensional
rest-mass density distribution (left panel) and the Lorentz factor
distribution (right panel) for model $\mathtt{A6}$ with $a/M=0.9$ and
$V_\infty=0.5$.  As evident from the iso-density contours, the
accretion in the upstream region, is essentially spherical accretion,
while a very well defined shock cone, characterized by a strong
density gradient, forms in the downstream region. The shock cone,
where the accretion rate is the largest, is clearly distorted by the
spin of the black hole and its opening angle slightly decreases with
increasing distance from the center. We have also found that spinning
black holes have maximum rest-mass density in the shock cone which is
higher than for non spinning black holes. The right panel of
Fig.~\ref{fig_2D_stationary}, on the other hand, shows that the
overall flow is only very mildly relativistic, with a maximum Lorentz
factor $W_{\rm {max}}\lesssim 3$, even ahead of the shock front in the
vicinity of the black hole. It is also interesting to note that
accretion in the shock cone takes place with a very small Lorentz
factor. Matter inside the shock cone can accrete onto the black hole
only for radii smaller than a given {\em accretion radius} $r_a$,
defined as the radius where the accretion rate (as a function of
radius) becomes negative. The expression $r_a\approx
2M/(V_\infty^2+c^2_{s,\infty})$, which is valid in the case of a
purely-spherical accretion, provides a good estimate of the accretion
radius in the Bondi-Hoyle accretion. As an example, for a model
with $V_\infty=0.5$ and $c_{s,\infty}=0.1$, we found $r_a=6.1$, while
the predicted one is $r_a=7.6$.

To complement the information in Fig.~\ref{fig_2D_stationary},
Fig.~\ref{fig_shock_location} shows one-dimensional profiles at time
$t=10000\,M$ of the rest-mass density at different radial shells for
an asymptotic flow velocity $V_{\infty}=0.5$. More specifically, the
left panel refers to Schwarzschild black hole, while the right one to
a rotating black hole with spin $a/M=0.9$. Clearly, a sharp transition
in the density exists at the border of the shock cone and this is
particularly evident for the nonrotating black hole. Note also the
amount of distortion which is present in the case of a rotating black
hole; although this is in part due to our choice of Boyer-Lindquist
coordinates, similar distortions are present also when using different
and better suited coordinates, such as Kerr-Schild [see the discussion
  in~\citet{Font1999b}]. As already mentioned above, the angular
location of the shock is only weakly dependent on the radial distance,
with the size of the shock cone reducing only slightly when moving
away from the black hole. As a final remark we note that the large
degree of symmetry shown in the left panel of
Fig.~\ref{fig_shock_location} provides a convincing evidence of the
abilities of the code to maintain the symmetry across $\phi=0$.

%------------------------------------------------
\subsubsection{The flip-flop instability}
\label{The_flip-flop_instability}

As shown through numerical simulations performed by several authors
over the years in Newtonian physics, the Bondi-Hoyle accretion flow is
subject to the so called {\em flip-flop} instability, namely an
instability of the shock cone to tangential velocities, which
manifests in the oscillation of the shock cone from one side of the
accretor to the other. Such instability was first discovered by
\citet{Matsuda1987} in 2-D Newtonian axisymmetric simulations, and
later confirmed and further investigated by~\citet{Fryxell1988},
\citet{Sawada1989},~\citet{Benensohn1997},~\citet{Foglizzo1999},
\citet{Pogorelov2000}. Three dimensional simulations were performed
by~\citet{Ishii1993} and~\citet{Ruffert1997}, who confirmed the
occurrence of the instability, although with deformations of the shock
cone only very close the accretor. In spite of all these
investigations, however, the nature and the physical origin of the
flip-flop instability remain obscure.  According to
\citet{Foglizzo1999}, for example, local instabilities, such as the
Rayleigh-Taylor or the Kelvin-Helmholtz instabilities, should not
play a significant role and cannot account for the flip-flop
instability. On the other hand,~\citet{Soker1990} showed, through a
WKB analysis, that the two-dimensional axisymmetric accretion flow
(not the one considered in our work) can be unstable against
tangential modes, as well as against radial modes. When extending
these results to three-dimensional simulations, however, the
persistence of such instabilities is not fully clarified. 
Overall, it is fair to say that, although the flip-flop
instability has been confirmed numerically both in two and
three-dimensional simulations, it is still unclear whether the
physical mechanisms driving the instabilities in the two cases are the
same or not. Finally, to further complicate the scenario, no evidence
was found by~\citet{Font1998g} of such instability within a
relativistic framework, although the authors argued that an
instability might develop for very large values of the asymptotic Mach
number.

More recently,~\citet{Foglizzo2005} performed a very detailed analysis
of the unstable behavior of Bondi-Hoyle accretion flows, suggesting
that the instability may be of advective-acoustic nature, both in the
case of shocks attached or detached (bow shocks) to the accretor.
Moreover, they also stressed that, though physical, the instability is
likely to be triggered or dumped by numerical effects such as the
carbuncle phenomenon at the shock, the boundary condition at the
surface of the accretor and the grid size.

Although a detailed analysis of the flip-flop instability is beyond
the scope of this paper, which is rather focused on the development
of QPOs in Bondi-Hoyle accretion flows, our simulations confirm for
the first time the occurrence of the instability even in a
relativistic regime. Fig.~\ref{flip-flop}, for instance, reports four
snapshots of the rest-mass density in a long term evolution of a model
when the instability is fully developed. The shock cone in the
downstream region oscillates back and forth in the orbital plane in
this representative model with black hole spin $a/M=0.5$ and
$V_{\infty}=0.4$. The reason why such unstable behaviour was not
noticed by~\citet{Font1999b} may be due to the short term evolution
that they were forced to consider at the time, or, as suggested
by~\citet{Foglizzo2005}, because of the high values of $2M/r_a=
(V_\infty^2+c^2_{s,\infty})$ that they considered.

\begin{figure}
\psfig{file=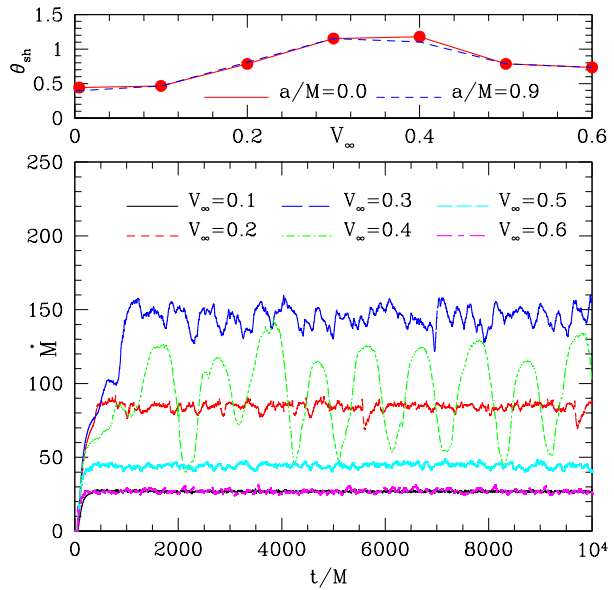,width=8.2cm}
\caption{\textit{Top panel:} shock opening angle in radians as
  computed at $r=4.43M$, $t=10000\,M$, as a function of
  $V_{\infty}$. Shown with different lines are the values for a
  Schwarzschild black hole (red solid line) and for a Kerr black hole
  (blue dashed line). \textit{Bottom panel:} time evolution of the
  mass accretion rate, computed at $r=6.08M$, for different value of
  $V_{\infty}$ and a rotating black hole with spin $a/M=0.9$. The mass
  accretion rate manifests high amplitude oscillation in the two
  flip-flop unstable models with $V_{\infty}\sim 0.3-0.4$.}
\label{opening_angle}
\end{figure}

Fig.~\ref{opening_angle} shows additional key features of the
flip-flop instability. The top panel, in particular, reports the shock
opening angle, as a function of the asymptotic velocity, computed in
radians at $r=4.43\,M$ and $t=10000\,M$.  Shown with different lines
are the values for a Schwarzschild black hole (red solid line) and for
a Kerr black hole (blue dashed line).  The points with
$V_{\infty}=0.3$ and $V_{\infty}=0.4$ correspond to models which are
flip-flop unstable and they are located close to the local maximum of
the shock opening angle. This results confirms what already reported
by~\citet{Livio1991}, namely that the instability manifests when the
shock opening angle is larger than a given threshold. Once the
instability is fully developed, the mass accretion rate experiences
high amplitude oscillations (for any value of the spin
considered). This is shown in the bottom panel of
Fig.~\ref{opening_angle}, reporting the evolution of the
one-dimensional mass accretion rate $\dot{M}$ defined as
\begin{equation}
\dot{M}\equiv -\int_0^{2\pi}{\alpha \sqrt{\gamma} \rho W\left(V^r
    - \frac{\beta^r}{\alpha}\right)d\phi}\,,
\end{equation}
for different choices of $V_{\infty}$; the specific values reported in
the figure refer to a shell at $r=6.08\,M$ and a rotating black hole
with spin $a/M=0.9$. It is therefore evident that for small values of
$V_{\infty}$, both the mass accretion rate and the shock opening angle
are increasing functions of $V_{\infty}$. However, the accretion is no
longer efficient above a critical value of $V_{\infty}\simeq 0.4$, and
any further increase of $V_{\infty}$ causes both the accretion rate
and the maximum rest-mass density in the shock cone to decrease.

%

%###############################################################################
%*****************************************************************************
%*****************************************************************************

\subsection{QPOs in the shock cone}
\label{QPOs_in_the_shock_cone}

\begin{figure*}
\psfig{file=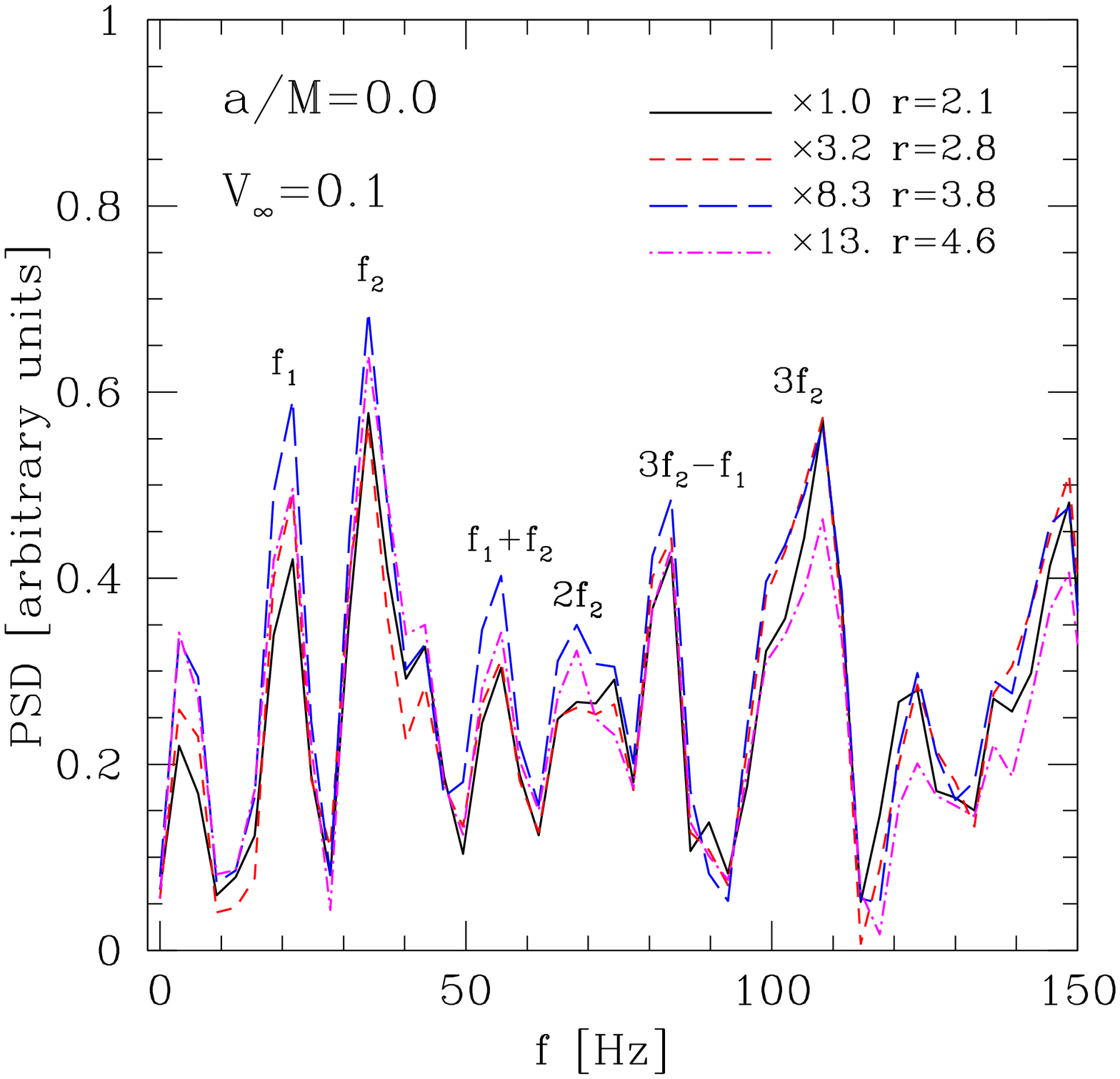,width=7.0cm}
\hskip 1.0cm
\psfig{file=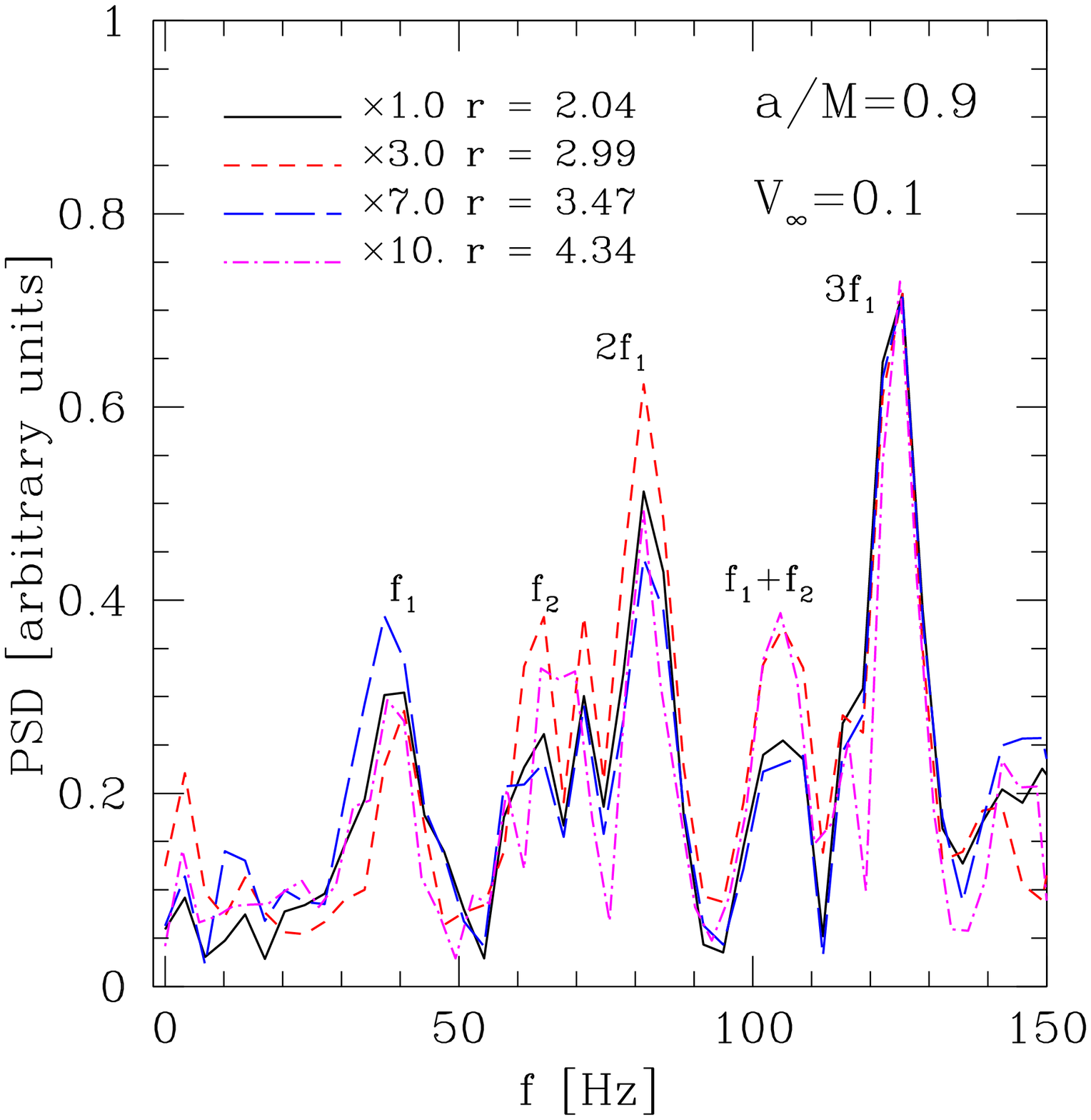,width=7.0cm}
\caption{Power spectra of the timeseries of the rest-mass density as
  computed at the center of shock cone for a flow with asymptotic
  velocity $V_{\infty}=0.1$. The left and right panels refer to black
  holes with spins for $a/M=0.0$ and $a/M=0.9$, respectively, while
  the mass is the same in the two cases and $M=10\,M_{\odot}$. The
  different curves refer to timeseries recorded at different radial
  positions but have the same azimuthal position.}
\label{fig_power_spectra}
\end{figure*}

As mentioned in the Introduction, one of the most important results
presented in this work is about the development of QPOs in the shock
cone that forms in the downstream region once the system has
relaxed to the stationary state. 
Note that because a 
  stationary state is necessary for the development of the QPOs, the
  latter are not found when the flip-flop instability
  develops. 
Although the occurrence of QPOs in Bondi-Hoyle accretion has
not been discussed before in previous numerical investigations, Fig.~8
of~\citet{Font98c} does shows a clear periodic behavior of the
accretion rate. So, although we can claim to be the first ones to have
pointed out the appearance of this effect and to have investigated it
in detail, we cannot claim to be the first to have shown found it from
numerical simulations.

In order to extract as much information as possible about the emerged
phenomenology and provide a first physical explanation, we have
carried out an extensive Fourier analysis of the different dynamical
quantities. The two panels of Fig.~\ref{fig_power_spectra}, for
instance, show the power spectra obtained from the timeseries of the
rest-mass density as computed in a region which is in the middle of
the shock cone for a flow with asymptotic velocity
$V_{\infty}=0.1$. The left and right panels refer to black holes with
spins for $a/M=0.0$ and $a/M=0.9$, respectively, while the mass is the
same in the two cases and equal to $M=10\,M_{\odot}$. The different
curves refer to timeseries recorded at different radial positions but
have the same azimuthal position. After performing a series of tests
at different resolutions we have estimated the error bar in the
measure of each frequency to be $\sim 3 {\rm Hz}$.

A number of comments should be made regarding the spectra
reported in
Fig.~\ref{fig_power_spectra}.
The first one is that the modes
of oscillation are essentially independent of the radial position, \ie
the power spectra at different radii overlap extremely well. This is a
clear indication that the modes are local waves but they are global
eigenmodes of the system. 

The second comment is that, among the modes reported in
Fig.~\ref{fig_power_spectra}, some are genuine eigenmodes, while
others are simply the result of nonlinear couplings. In particular,
the modes with frequencies $f_1=21 \,{\rm Hz}$ and $f_2=34 \,{\rm Hz}$
in the left panel of Fig.~\ref{fig_power_spectra} are genuine
eigenmodes, while those at $55, 68, 83$ and $106 \,{\rm Hz}$ are
given, within a few percent error, by nonlinear couplings of the
eigenmodes, \ie $f_1+f_2$, $2f_2$, $3f_2-f_1$, and $3f_2$,
respectively. Similarly, the modes $f_1=40 \ {\rm Hz}$ and $f_2=64
\,{\rm Hz}$ in the right panel of Fig.~\ref{fig_power_spectra} are
genuine eigenmodes, while the modes at $81, 104$ and $124 \,{\rm Hz}$
are given, within few percent errors, by $2f_1$, $f_1+f_2$, and
$3f_1$, respectively. We recall that this behavior is typical of
physical systems governed by non linear equations in the limit of
small oscillations \citep{Landau-Lifshitz1} and has already been
pointed out by~\citet{Zanotti05} in the context of oscillation modes
in thick accretion discs around black holes. It should also be
remarked that, while a large number of modes due to nonlinear coupling
exists, the identification becomes difficult for frequencies larger
than $100\,{\rm Hz}$. Similarly, the low-frequency mode at $\sim
5\,{\rm Hz}$, which is visible in both the panels of
Fig.~\ref{fig_power_spectra}, could be a genuine mode but it appears
with much smaller intensity and not all radii. As a result, its
classification as a genuine mode will require more accurate
simulations and on much longer timescales.

\begin{figure}
\vspace{0.5cm}
\psfig{file=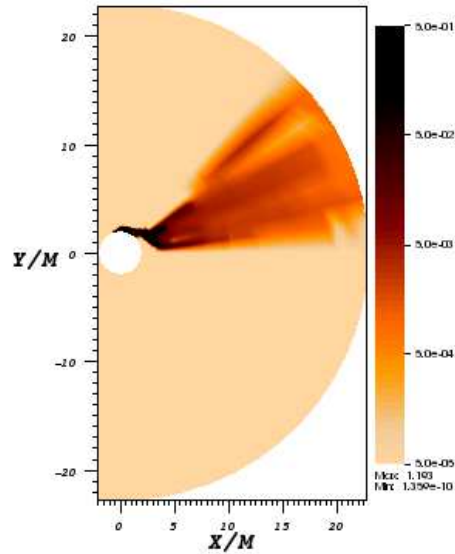,width=8.2cm}
\caption{Spatial distribution of the power spectral density at $f=124
  {\rm Hz}$ for a black hole of mass $M=10\,M_{\odot}$ and spin
  $a/M=0.9$ (\cf right panel of Fig.~\ref{fig_power_spectra}). A very
  similar behaviour would be shown in the case of a nonrotating black
  hole.}
\label{QPO_Inten}
\end{figure}

The global nature of the oscillation modes inside the shock cone is
clearly shown in Fig.~\ref{QPO_Inten}, which represents the power
spectral intensity of the most powerful mode in the right panel of
Fig.~\ref{fig_power_spectra}, namely of the mode at $f=124\,{\rm Hz}$
(A very similar behaviour can be seen in the case of a nonrotating
black hole.). In other words, for each grid-cell inside the shock cone
we store the evolution of the rest-mass density and compute from this
the power spectral density. We then consider a single frequency and
study how its intensity is distributed (in arbitrary units) inside the
shock cone.

Note that the intensity is computed over the whole computational
domain, but it has a non-negligible amplitude only inside the shock
cone, and this amplitude becomes increasingly stronger near the black
hole (We recall that in general we do not expect the amplitude to be
constant in the shock cone, as this depends on the specific
eigenfunction of the mode). What shown in Fig.~\ref{QPO_Inten} is not
specific to the mode at $f=124\,{\rm Hz}$ and we have verified that
the spatial distribution of the intensity manifests a similar pattern
for all of the other modes. Besides providing evidence that these are
indeed global oscillation modes of the flow, Fig.~\ref{QPO_Inten} also
highlights that QPOs can be excited in the shock cone of a Bondi-Hoyle
type accretion and that these are particularly stronger close to the
black hole.

Interestingly, when the flip-flop instability is triggered and
develops, the power spectra inside the shock cone change considerably
and in these cases only the periodicity of the shock-cone
oscillation can be found,
which is then
accompanied by the usual nonlinear couplings. In general, therefore,
the effect of the flip-flop instability is that of suppressing most of
the internal modes of oscillations. A more detailed analysis of this
process, as well as of the flip-flop instability will be presented in a
future work.

\begin{figure*}
\psfig{file=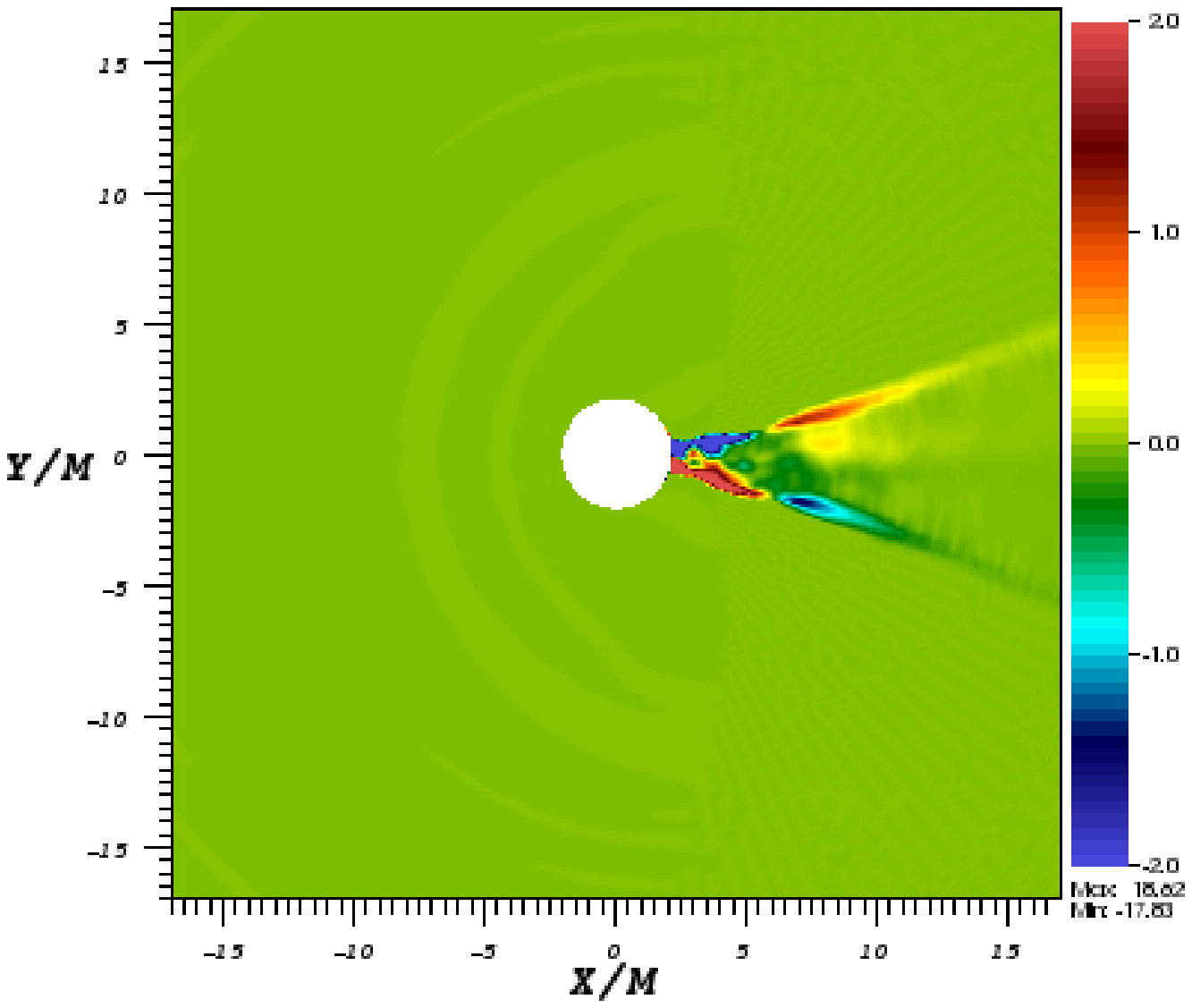,width=7.0cm}
\hskip 1.0cm
\psfig{file=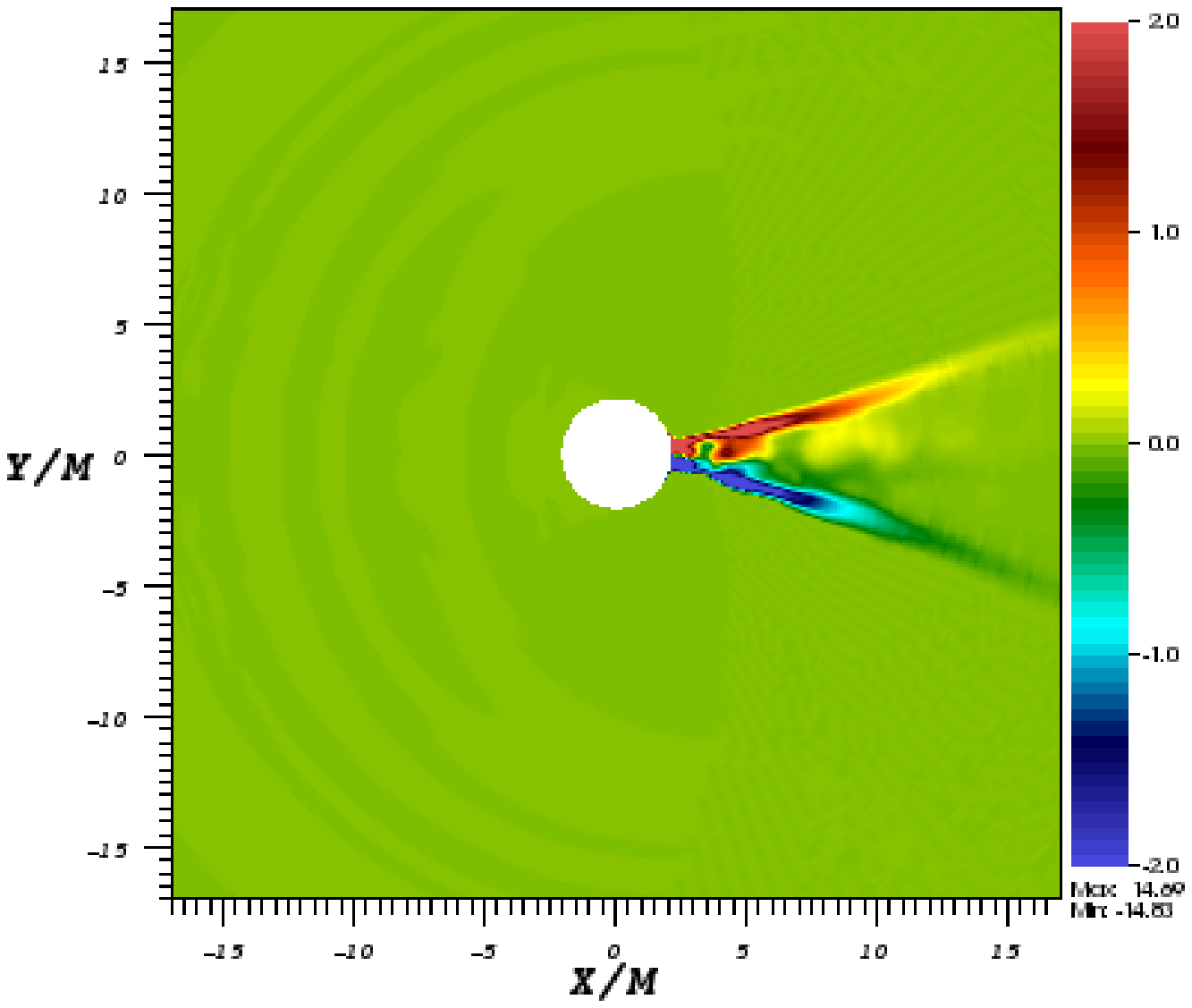,width=7.0cm}
\caption{Numerical eigenfunctions computed as $\rho(r,\phi)-{\tilde
    \rho}_0(r,\phi)$, where ${\tilde \rho}_0(r,\phi)$ is the density
  distribution when the flow is stationary (\ie for $t\gtrsim 1500\,M$).  
The left and the right panels report two
  snapshots at $t=1871M$ and $t=2162M$, and are
  separated by half a period of oscillation; both refer for to a black
  hole with spin $a/M=0.0$ and $V_{\infty}=0.1$.
\label{fig_2D_eigenfunction}}
\end{figure*}
 A linear perturbative analysis of the stationary pattern would
certainly be of great interest for three important reasons.
Firstly, in order to identify unambiguously 
which of the modes are genuine (or fundamental) and which
are the effect of non-linear coupling. 
Secondly,
to compute the expected
eigenfrequencies and eigenfunctions of the oscillations we have
discovered numerically. Finally 
to clarify why some modes are more efficiently excited than others. 
However, such perturbative analysis
is probably going to be extremely complicated, mainly because of the
absence of an analytic solution for the stationary flow inside the
shock cone. To counter these limitations and perform a first
rudimentary analysis of the eigenfunctions, we report in
Fig.~\ref{fig_2D_eigenfunction} a map of what can be regarded as a
numerical eigenfunction for the model $\mathtt{C2}$ ($V_{\infty}=0.1$
and spin $a/M=0.0$). In fact, the two panels of
Fig.~\ref{fig_2D_eigenfunction} show the difference
$\rho(r,\phi)-{\tilde \rho}_0(r,\phi)$, where ${\tilde
  \rho}_0(r,\phi)$ is the density distribution when the flow is
stationary (\ie for $t\gtrsim 1500\,M$). Considering, for example, the
left panel, which corresponds to $t=t_1=1871M$ and using Cartesian
coordinates, we show in the upper part of the figure, \ie for $y/M>0$,
the density perturbation along the shock is negative close to the
black hole, while it is positive far from it. A specular behavior is
detected in the lower part of the figure where $y/M<0$. The computed
perturbation oscillates in time with a given period $P$, and the right
panel of the figure shows a snapshot at time $t_2=t_1+P/2=2162M$, \ie
at half a period of oscillation. Although heuristic, this procedure
allows to compute the period of oscillation of the perturbation with a
very good accuracy. The corresponding frequency, again computed
assuming $M=10M_{\odot}$, is, in fact, $f\sim 35 \,{\rm Hz}$ and it
matches very well with the most powerful mode, that with frequency
$f_2$, reported in the left panel of Fig.~\ref{fig_power_spectra}. To
highlight the dependence of the eigenfrequencies on the black hole
spin, we report in Table~\ref{table:spin_frequency} their values as
measured within the shock cone from the evolution of the rest-mass
density. The mass
of the black hole is assumed to be $M=10\,M_{\odot}$, but the
frequencies can be easily computed for an arbitrary mass since they
scale linearly with the black hole mass. The data in
Table~\ref{table:spin_frequency} suggests that a linear scaling is
present also for the black hole spin, but clearly more data is necessary to
confirm this scaling, which we will investigate in a future work.

As a final remark we note that, overall, the phenomenology we have
found show strong similarities with that reported for perturbed
relativistic tori orbiting around a black hole \citep{Zanotti03}. In
that case it was shown by~\citet{Rezzolla_qpo_03b} through a
perturbative investigation that the eigenfunctions and
eigenfrequencies were those corresponding to the $p$ modes of the
torus. Such modes, also known as inertial-acoustic waves, are closely
related to the propagation of sound waves in the perturbed fluid and
have pressure gradients and centrifugal forces as the main restoring
forces. The major difference of the physical conditions investigated
by~\citet{Rezzolla_qpo_03b} with the ones considered here is that
centrifugal forces play a negligible role in Bondi-Hoyle
accretion flows. Apart from this, however, the physical nature of the
modes is essentially the same in the two cases. When applied to the
oscillations of a thick disc around a compact object, such modes
motivated the proposal of a model for explaining the detection of High
Frequency Quasi Periodic Oscillations (HFQPOs) in the spectra of X-ray
binaries~\citep{Rezzolla_qpo_03a}.
{\begin{table}
  \caption{Frequencies of the first two genuine modes measured within
    the shock cone from the evolution of the rest-mass density. The
    values refer to different black hole spins and different
    asymptotic velocities, 
    while
    the mass of the black hole is always assumed to be
    $M=10\,M_{\odot}$. 
    Note that the frequencies scale linearly with $M$ and the values
    reported have an error bar of $\sim 3$ Hz. 
 \label{table:spin_frequency}}
\begin{center}
  \begin{tabular}{cccc}
    \hline \hline
    $a/M$ & $V_{\infty}$ & $f_1$ [Hz]& $f_2$  [Hz]\\
    \hline
    $0.0$ &  $0.1$ &  $21$ & $34$ \\
    $0.5$ &  $0.1$ &  $32$ & $53$ \\
    $0.9$ &  $0.1$ &  $40$ & $64$ \\
    \hline
    $0.0$ &  $0.2$ &  $32$ & $46$ \\
    $0.9$ &  $0.2$ &  $46$ & $84$ \\
    \hline
    \hline
  \end{tabular}
\end{center}
\end{table}

%****************************************************************************

\section{Astrophysical applications}
\label{Astrophysical_applications}

\subsection{The case of ${\rm Sgr~A}^\ast$}

A possible application of the results discussed in the previous
Section and, in particular, of the development of QPOs in the
downstream part of a Bondi-Hoyle flow, is offered by the source
Sagittarius A$^\ast$ (or ${\rm Sgr~A}^\ast$). Such source is located at
the centre of our Galaxy and is widely believed to be a supermassive
black hole with an estimated mass of $\sim 4.1 \pm 0.6 \times 10^6 \,
M_{\odot}$~\citep{Ghez:2008,Gillessen:2009}. ${\rm Sgr~A}^\ast$ is one
of the most intensely observed astronomical objects and the
phenomenology of its emission is both rich and particularly
complex. Interestingly, however, QPOs from its emission have been
observed by~\citet{Aschenbach2004} (see also~\citet{Aschenbach2009}),
when analyzing two distinct near-infrared and X-ray
flares~\citep{Genzel2003}. In particular,~\citet{Aschenbach2004}
reported five different peaks at periods of $100$ s, $219$ s, $700$ s,
$1150$ s, and $2250$ s in the power spectral density of such
flares. Soon after,~\citet{Abramowicz2004} noticed that
$(1/700):(1/1150):(1/2250)\approx 3:2:1$, that is, the periods are in
a harmonic ratio. Indeed, while discussing the same
source,~\citet{Torok2005} confirmed the measurement of a double peak
of QPOs in $\rm {Sgr~A}^\ast$ in the same typical $3:2$ ratio observed
in several low mass X-ray binaries and suggested the epicyclic
resonance model by~\citet{Abramowicz2003} as a possible explanation
for this phenomenology. ~\citet{Chan2009}, on the other hand,
performed two dimensional magnetohydrodynamical simulations of a
putative accretion disc in $\rm{Sgr~A}^\ast$ and showed that the QPOs
frequencies could be due to non-axisymmetric density perturbations,
and which may be used to set a lower limit on the orbital period at
the innermost stable circular orbit.

In addition to QPOs, $\rm {Sgr~A}^\ast$ is likely to have a
Bondi-Hoyle accretion flow. This possibility was first pointed out
by~\citet{Melia1992}, who, after looking at the broad He I, Br$\alpha$
and Br$\gamma$ emission lines, inferred that there is a strong
circumnuclear wind near the dynamical center of the Galaxy. In this
scenario, the central black hole is assumed to be fed by stellar winds
within several arcseconds from the compact
object~\citep{Shcherbakov2009}. In order to investigate this
possibility further,~\citet{Ruffert1994} performed the first
three-dimensional hydrodynamical simulations of this system and
computed the line-integrated flux assuming that the emission is
dominated by bremsstrahlung. By collecting information coming from
more recent observations,~\citet{Cuadra2006a} and~\citet{Cuadra2006b}
performed SPH simulations of wind accretion onto $\rm {Sgr~A}^\ast$,
including optically thin radiative cooling and finding, among other
results, that most of the accreted gas is hot and has a
nearly-stationary accretion rate. Although the observed luminosity is
some orders of magnitude smaller than what is predicted by the
Bondi-Hoyle model, this may be due to a low radiative efficiency of
the accretion flow, rather than to a failure of the Bondi-Hoyle model
itself.

Should the two observational features mentioned above be confirmed by
additional and more accurate observations, there would be a unique
astronomical realization of a source for which both QPOs and
Bondi-Hoyle accretion are simultaneously present. It should be
stressed that the typical lengthscale at which the QPOs develop in
our simulations is $\approx 20\,M$, which thus corresponds to
$\approx 4\times10^{-6}\,{\rm pc}$ for the estimated mass of ${\rm
Sgr~A}^\ast$. This lengthscale is much smaller than the 
closest approach to the Galactic Center of the closest star S16, which is
around $\approx 2\times10^{-4} {\rm pc}$, as reported by
\citet{Ghez2005}. 
Therefore, the Bondi-Hoyle accretion flow that
could potentially be present in ${\rm Sgr~A}^\ast$, would take take
place in a gas-dominated region very close to the central black hole
and far from the orbits of the S-stars.

Interestingly, after analyzing recent data about a high-level X-ray
activity of ${\rm Sgr~A}^\ast$ observed with
XMM-Newton~\citep{Porquet2008, Aschenbach2009}, \citet{Aschenbach2009} reports nine prominent
peaks, named $\nu_1\ldots \nu_9$.  Out of them, only three are
identified as fundamental by~\citet{Aschenbach2009}, namely $\nu_6$,
$\nu_7$ and $\nu_9$, while the other six are obtained as linear
combinations of the fundamental ones, like, for instance,
$\nu_6+\nu_7$, $\nu_6+\nu_9$, $\nu_7-\nu_9$.  \citet{Aschenbach2009}
interprets the three fundamental modes in terms of orbital, radial
epicyclic and vertical epicyclic frequencies.  In doing so, a
specific radius where the oscillation is excited must be provided,
the above osscillation frequencies being intrinsecally local. On the
contrary, no radial specification is required within our
interpretation, since the modes are global and just confined within
the shock cone.  Finally, although \citet{Aschenbach2009} does not give
any explanation for the other six modes, they are naturally
provided within our interpretation of a Bondi-Hoyle accretion in
which the downwind shock cone acts as a cavity, generating a whole
series of nonlinear couplings among few genuine and trapped pressure
modes.

%--------------------------------------------
\subsection{QPOs in HMXBs} 

A second application of our result is in principle represented by QPOs
observed in the spectra of high mass X-ray binaries (HMXBs), which are
composed of a compact object and of an early type (OB) star. The
catalog by~\citet{Liu2006} lists 114 HMXBs candidates in the Galaxy,
some of which are believed to contain a black hole, like Cyg X-1, NGC
5204~\citep{Liu2004}, M 33 X-7~\citep{Pietsch2004}, M 101
ULX-1~\citep{Mukai2005}. Because in these systems the accretion flow
occurs preferentially in the form of an accretion wind rather than in
the form of an accretion disc, the Bondi-Hoyle accretion flow has been
traditionally considered very relevant for them (see the review
by~\citet{Edgar2004} for the application of the Bondi-Hoyle solution
to binary systems). However, the QPOs that have been detected in HMXBs
have frequencies that are seen to lie in the range of $1$ mHz to $400$
mHz, with the remarkable exception of XTE J0111.2-7317, which shows a
QPO feature at $1.27$ \Hz \citep{Kaur2007}. On the other hand, the
QPOs we have computed assuming a black hole of mass $M=10\,M_{\odot}$ have
frequencies which lie in the range from $1$ \Hz to $500$ \Hz, and
therefore that overlap only marginally with the observed ones.

%###############################################################################

\section{Conclusions}
\label{Discussion_and_conclusions}

Although viscous accretion discs represent the most natural channel by
means of which compact objects accrete large amounts of matter with
significant angular momentum, non-spherical accretion flows appear in
all those situations in which the accreting matter has only a modest
amount of angular momentum, such as in winds. Bondi-Hoyle accretion is
the most representative example of this type of flow and its study via
numerical investigation has a long history, both in Newtonian and in
general-relativistic physics. We have reconsidered this old problem by
performing new two-dimensional and general-relativistic simulations
onto a rotating black hole. Besides recovering many of the features of
this flow which were discussed by other authors over the years, we
have also pointed out a novel feature. More specifically, we have
shown that under rather generic conditions a shock cone develops in
the downstream region of the flow and that such a cone acts like a
cavity trapping pressure modes and giving rise to QPOs.

These modes are global in the sense that they represent harmonic
oscillations across the cavity, but have amplitudes that are larger in
the region very close to the black hole, \ie for $r \lesssim 10
M$. While the black hole spin influences the absolute frequencies of
the trapped modes, which lie in the range from $1$ \Hz to $500$ \Hz
for a representative black hole of mass $M=10\,M_{\odot}$, it does not
affect the fact that they appear in a series of integer numbers
$1:2:3$. This is because nonlinear coupling of modes is common in systems
governed by nonlinear equations, and once a mode is excited, all of its
integer multiples are also excited, thus producing a wide range of
possible ratios of integer numbers.

In addition to pointing out this novel feature of the Bondi-Hoyle
accretion, we have discussed its possible application to the
phenomenology reported in ${\rm Sgr~A}^\ast$, where both
QPOs~\citep{Aschenbach2004} and Bondi-Hoyle
accretion~\citep{Melia1992} are likely to be present. Remarkably, a
large family of linear combinations of modes has been identified
recently by~\citet{Aschenbach2009} after analyzing data about a
high-level X-ray activity of ${\rm Sgr~A}^\ast$ observed with
XMM-Newton~\citep{Porquet2008}. This phenomenology could indeed be
interpreted in terms of a shock cone that behaves like a cavity and
that generates a whole class of nonlinear coupling among pressure
modes.

Finally, we have provided the first evidence for the occurrence in a
general relativistic context of the so called flip-flop instability of
a Bondi-Hoyle flow. More specifically, we have shown that for a fixed
choice of the black hole spin and of the sound speed, there exists a
critical value of the asymptotic flow velocity at which the shock cone
undergoes large-scale and coherent oscillations. When this happens,
the opening angle of the shock cone reaches its maximum value, the
accretion rate increases considerably and is no longer stationary.  A
more comprehensive analysis is needed to clarify the physical nature
of the instability and its dynamics in the relativistic regime. This
will be the focus of a future work.

Finally, among the future improvements of our work we plan to
investigate the effects of a magnetic field, which is certainly
likely to play a role but whose effective contribution has not been
considered yet in numerical simulations of Bondi-Hoyle accretion
flows.  In particular, the potential interplay of the
magnetorotational instability (MRI) with the flip flop instability
has to taken into account, though it is not clear whether a local
instability such as the MRI can substantially affect a global
instability.  As far as the occurrence of QPOs, on the other hand,
we believe that it will remain essentially unchanged. The reason for
this is that as long as a shock cone forms in the downstream part of
the flow, QPOs will be naturally excited and trapped, the only
difference being that they will be of magnetosonic nature rather
than of a sonic one, and hence with eigenfrequencies that will
depend not only on the properties of the shock-cone (and thus of the
fluid) but also on the strength of the magnetic field.

\section*{Acknowledgments}

It is a pleasure to thank Jos\'e A. Font for
carefully reading the manuscript and for very helpful comments.
The numerical calculations were performed at the National Center for
High Performance Computing of Turkey (UYBHM) under grant number
10022007 and TUBITAK ULAKBIM, High-Performance and Grid-Computing
Center (TR-Grid e-Infrastructure). This work was supported in part by
the DFG grant SFB/Transregio~7.

\bibliographystyle{mn2e}

\bsp

\label{lastpage}

\end{document}